\documentclass{ws-procs9x6}

\begin{document}

\title{APPLICATIONS OF CONTINUUM SHELL MODEL}

\author{A. VOLYA}

\address{Department of Physics, Florida State University,\\
Tallahassee, FL 32306-4350, USA\\
E-mail: volya@phy.fsu.edu}

\begin{abstract}
The nuclear many-body problem at the limits of stability is
considered in the framework of the Continuum Shell Model that
allows a unified description of intrinsic structure and reactions.
Technical details behind the method are highlighted and practical
applications combining the reaction and structure pictures are
presented.
\end{abstract}


\bodymatter
\section{Introduction}

In this presentation we discuss specific features of the
Continuum Shell Model (CoSMo), the approach based on the
projection formalism \cite{feshbach58}, formulated in
the classical book \cite{mahaux69} and developed into a practical
instrument in Refs. \cite{rotter91,volya_PRC67}. The
whole problem of many-body physics on the verge of stability has
been extensively explored in the past, especially in relation to
weakly bound nuclei. Alternative formulations and their first
applications can be found, for example, in Refs.
\cite{betan02,michel02,okolowicz03}.

The goal of this paper is to highlight complimentary views on the
nuclear many-body physics from the ``inside" (structure) and
``outside" (reactions) perspectives. The structure view is based
on the traditional shell model where the effective Hamiltonian to
be diagonalized plays the central role. New contributions to the
effective Hamiltonian coming from the presence of continuum bring
in non-Hermiticity and energy dependence. Overcoming these
complications, it is possible to calculate in the same framework
the cross sections of reactions, with their energy dependence and
possible resonance behavior. The complementary picture that
appears from the side of nuclear reactions is important for
identifying resonances and comparison with experiment. While the
shell model approach to the many-body structure in discrete
spectrum is firmly established the many-body reaction physics is
usually left for more phenomenological tools of the reaction
practitioners. The purpose of Secs. \ref{unit} and \ref{cheb} in
this work is to accentuate on novel methods involved in
calculation of Green's functions and associated time evolution
operators that stay behind CoSMo.

The example of a realistic application presented in the last
section is a central point of the paper. The chain of helium
isotopes is shown where a single picture combines different
methods and different points of view on the same problem. The
bound states of the conventional shell model below threshold are
followed at higher energies by the solutions of the CoSMo
effective Hamiltonian revealing resonances that coincide with the
complex poles of the scattering matrix. The same resonances appear
in the neutron scattering cross section plotted in the same
figure. The discrepancies between the cross section peaks and
resonance states emphasize subtle features of many-body dynamics
in a marginally stable system.

\section{Structure\label{structure}}

Using the projection formalism one can eliminate the part of the
Hilbert space related to particle(s) in continuum. This results in
the effective Hamiltonian ${\cal H}$ that acts only in the
``intrinsic'' shell model space,
\begin{equation}
{\cal H}(E)=H_0+\Delta(E)-\frac{i}{2}W(E).           \label{1}
\end{equation}
Here the full Hamiltonian $H_0$ is restricted to intrinsic space, and 
is supplemented with
the Hermitian term $\Delta(E)$ that describes virtual particle
excitations into excluded space and the imaginary term $W(E)$
representing irreversible decays to the continuum. The new parts
of the Hamiltonian (\ref{1}) are found in terms of the matrix
elements of the full original Hamiltonian that link the internal
states $|1\rangle$ with the energy-labeled external states
$|c;E\rangle$: $A^c_1(E)=\langle 1|H_{0}|c; E\rangle,  $
\begin{equation}
\Delta_{12}(E)={\rm P.v.}\,\int dE'\sum_{c}\frac{A^{c}_{1}(E')
A^{c\ast}_{2}(E')}{E-E'},      
\,\,
W_{12}(E)=2\pi\sum_{c({\rm open})}A^{c}_{1}A^{c\ast}_{2},
                  \label{3}
\end{equation}

Reduction of the effective space does not go without a price.
The new properties of the effective Hamiltonian (\ref{1}) are:\\
1. For the description of unbound states the effective Hamiltonian
is non-Hermitian which reflects the possible leak of probability
from the internal system. \\
2. The Hamiltonian has explicit energy dependence, making the
internal dynamics highly non-linear.\\
3. The additional terms in the Hamiltonian that appear as a result
of projection can be complicated. Even with exclusively two-body
forces in the full space, the many-body interactions appear in
the projected effective Hamiltonian. 

By construction, the eigenvalue problem
\begin{equation}
{\cal H}(E)|\alpha\rangle={\cal E} |\alpha \rangle \label{5}
\end{equation}
determines the internal part of the solution which is subject to
the regular boundary condition inside matched to purely outgoing
waves in the continuum. For energies $E$ below all thresholds, the
amplitudes $A^{c}_1(E)$ vanish, and Eq. (\ref{5}) determines
discrete bound states with real ${\cal E}=E$. Above decay
thresholds, Eq. (\ref{5}) has no real energy solutions, and the
stationary state boundary condition can not be satisfied. The
similarity of this problem to that for the bound states makes it
appealing to depart from the real axis and to find discrete
non-Hermitian eigenvalues. The complex energy roots ${\cal E}$ of
(\ref{5}) correspond to poles of the scattering matrix, see
discussion below, and represent the many-body resonant Siegert
states \cite{siegert39}.

The transition into a complex energy plane may be rather
impractical, it causes computational complications related to
numerous branch cuts and unphysical roots, the relation to
observables becomes complicated and rather remote. As an
alternative, the Breit-Wigner approach \cite{breit36} is commonly
used. Here the resonances ${\cal E}=E-(i/2)\Gamma$ are defined as
$
{\rm Re}\,[{\cal E}_\alpha(E)]=E$ and $ \Gamma_{\alpha}=-2{\rm
Im}\, [{\cal E}_{\alpha}(E)].                
$
In the limit of a small imaginary part (narrow resonances)
various definitions are equivalent. In the application of the
CoSMo discussed below we use the Breit-Wigner approach. However,
the general difficulty in parameterizing resonances in terms of
centroid energies and widths should be noted. As demonstrated in
Sec. {\ref{he}}, the problem becomes especially acute for broad
resonances, high density of states, or in near-threshold
situations. A look at the problem from the observable cross
sections is imperative.

\section{Reactions}

The picture where the nuclear system is probed from ``outside" is given
by the transition matrix defined within the general scattering theory
\cite{mahaux69},
\begin{equation}
T^{ab}(E) = \sum_{12}A^{a\ast}_{1}(E)\, \left(\frac{1}{E-{\cal
H}(E)}\right)_{12}A^{b}_{2}(E),                 \label{7}
\end{equation}
\begin{figure}[ht]
\vskip -0.4 cm
\begin{minipage}{2.4 in}
\caption{Reaction process: the entrance channel $b$ with amplitude
$A^{b}_2$ continues through internal propagation started in the
intrinsic state $|2\rangle$ driven by the non-Hermitian
energy-dependent effective Hamiltonian (\ref{1}) (with all
excursions into continuum space included), and ends by exit from
the intrinsic state $|1\rangle$ into the channel $a$ as described
by the amplitude $A^{a\ast}_{1}(E)$.
 \label{Tmatrix}}
\end{minipage}
\begin{minipage}[1]{1.8 in}
\centerline{\epsfxsize=1.6in\epsfbox{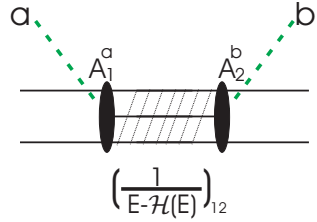}}
\end{minipage}
\vskip -0.4 cm
\end{figure}
The same transition amplitudes and propagation via intrinsic space
drive the process shown schematically in Fig. \ref{Tmatrix}.

The poles of the transition matrix and related full scattering
matrix $S=1-2\pi i T$ are the eigenvalues of Eq. (\ref{5}) located in
the lower part of the complex energy plane. The reaction theory is fully
consistent with resonant description in Sec.~\ref{structure}. 
However, complexity of
the many-body propagator in Eq. (\ref{7}) with numerous poles,
interfering paths and energy dependence can make the observable
cross section which is a projection of poles onto a real energy
axis quite different from a collection of individual resonance
peaks.

The solution of Eq. (\ref{5}) with the large scale many-body
Hamiltonian is a complicated task as extensively discussed in
Refs. \cite{volya_PRC67}. The calculation of the
transition matrix (\ref{7}) and of the cross section is yet
another technical problem. The direct approach involving matrix
inversion at all energies is extremely difficult and time
consuming given large dimensions involved. The sharp resonances
typically present in the spectrum make the process numerically
unstable and require dense energy sampling to achieve a reasonable
cross section curve. Absence of absolute numerical precision leads
to instability near stable states embedded in the continuum where
decays are prohibited by symmetry considerations. This problem is
particularly troublesome within the $m$-scheme shell model
approach. To overcome these difficulties, an alternative method
has been developed which is discussed below.

\subsection{Unitarity and $R$-matrix\label{unit}}

The transition matrix (\ref{7}) with the dimensionality equal to
the number of open channels can be written as $T={\bf A}^\dagger
{\cal G} {\bf A}$, where the full effective Green's function
${\cal G}(E)={1}/({E-{\cal H}})$
includes the loss of probability into all decay channels. The
factorized form of the non-Hermitian part $W=2\pi {\bf A} {\bf
A}^\dagger$ in Eq. (\ref{1}), where ${\bf A}$ represents a channel
matrix (a set of columns of vectors $A^c_1$ for each channel $c$)
is the key for unitarity of the $S$ matrix \cite{durand76}. As
shown in \cite{sokolov89}, the simple iteration of the Dyson
equation using the definitions ${\cal H}=H-i W /2$ 
and
$G=(E-H)^{-1}$ leads to the following transition and scattering
matrices
\begin{equation}
T=\frac{R}{1+i\pi R}\,,\quad S=\frac{1-i\pi R}{1+i\pi R},
                                            \label{9}
\end{equation}
where the matrix $R={\bf A}^\dagger G {\bf A}$ is analogous to the
$R$-matrix of standard reaction theory; it is based on the
Hermitian part of the Hamiltonian $H=H_0+\Delta$ and computed as a function of
energy using Chebyshev polynomial expansion in Sec. \ref{cheb}.

\subsection{Time evolution of the system and Green's function\label{cheb}}

The technique behind the Green's function calculation in the CoSMo
extends the idea suggested in \cite{ikegami02} where densities of
states in molecular systems were computed using the Chebyshev
polynomial expansion of the time-dependent evolution operator.

First, the finite Hamiltonian matrix is rescaled and shifted by a
constant so that the spectrum is mapped onto a generic energy
interval [-1,1] using $H\rightarrow(H-E_+)/{E_-}.$ The procedure
involves scaling $E_-$ and shifting $E_+$ parameters,
$E_\pm=(E_{\rm max}\pm E_{\rm min})/2$, that are determined by the
upper and lower edges of the original spectrum $E_{\rm max}$ and
$E_{\rm min}$, respectively. Even within the traditional Lanczos
diagonalization, the rescaling, although not required, is
useful for providing numerical stability. Given a trivial
nature of the rescaling procedure and its reversal, below we do
not introduce special notations for the rescaled Hamiltonian.

The energy representation of the retarded propagator is given by
the usual Fourier image of the evolution operator,
\begin{equation}
{G}(E)=\frac{1}{E-H}=-i \int_0^\infty dt\, \exp(iEt) \exp(-iHt),
                                               \label{10}
\end{equation}
where ${H}$ is the Hamiltonian operator with a negative-definite
infinitesimal imaginary part. The expansion factorizes the
evolution operator using the Chebyshev polynomials as follows:
\begin{equation}
\exp(-iH t)=\sum_{n=0}^\infty\, (-i)^n(2-\delta_{n 0})\, J_n(t)\,
T_n(H),                                   \label{11}
\end{equation}
where $J_n(t)$ is the usual Bessel function and the Chebyshev
polynomials are defined as
$T_n[\cos(\theta)]=\cos(n\theta). $
In comparison to the Taylor expansion or other methods evaluating
the Green's function, the Chebyshev polynomials provide a complete
set of orthogonal functions covering uniformly the interval
[-1,1]. Although individual states can be resolved, the procedure
is most effective when a significant energy region is involved,
namely for overlapping resonances. The asymptotic of the Bessel
functions assures convergence of the series. The ``angle
addition'' equations that follow from the definition of polynomials,
\begin{equation}
2T_n(x) T_m(x)=T_{n+m}(x)+T_{n-m}(x)\,,\quad n \ge m, \label{13}
\end{equation}
are useful for the successive evaluation of series of vectors
$|\lambda_n\rangle=T_n(H)|\lambda\rangle $ using the following
iterative procedure:
\begin{equation}
|\lambda_0\rangle =|\lambda\rangle,\quad |\lambda_1\rangle = H
|\lambda\rangle, \quad \text{and}\,\,\, |\lambda_{n+1}\rangle
=2H|\lambda_n\rangle-|\lambda_{n-1}\rangle.
                                                 \label{14}
\end{equation}

In the CoSMo approach the calculations of reactions are performed
using the Fast Fourier Transformation of Eq. (\ref{10}), where the
expectation value of the evolution operator in (\ref{11}) is
computed using iterative matrix-vector multiplications (\ref{14}).
In this way the $R$-matrix is computed which is then used to
determine the cross section (\ref{9}). The Chebyshev polynomials
are divergent in the complex plane; therefore only the Hermitian
part $H=H_0+\Delta$ corresponding to the $R$ matrix can be used
in the evolution operator (\ref{11}). Using a conservative
estimate it can be shown that $n$ iterations lead to the energy
resolution $4E_{-}/n$. Unlike the reorthogonalization problem in
the Lanczos algorithm, lack of numerical precision in successive
matrix vector multiplication does not leads to significant
deterioration of the result. $n=1024$ was typically used for CoSMo
calculations.
\begin{figure}[ht]
\vskip -0.4 cm
\begin{minipage}[1]{2.4 in}
\centerline{\epsfxsize=2.2in\epsfbox{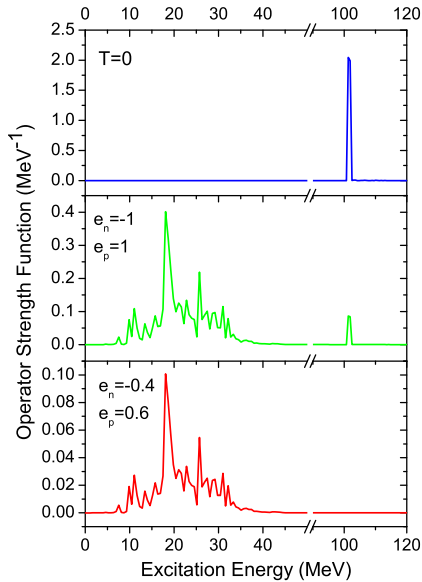}}
\end{minipage}
\begin{minipage}{1.8 in}
\caption{Strength function of a dipole operator. Upper plot:
effective charges for protons and neutrons are selected equal
leading to a pure isospin $T=0$ CM operator; only the CM states
with energies around 100 MeV have non-zero strength. Middle plot:
effective charges are selected as -1 and +1 for neutrons and
protons, respectively, the resulting mixed operator shows strength
in both CM and non-CM states. Lower plot: the effective charges
are selected as $e_n=-N/A=-0.4$ and $e_p=Z/A=0.6$, which for
$^{20}$O excludes CM component from the dipole operator; the
resulting strength shows no CM excitation.  \label{CMO20}}
\end{minipage}
\vskip -0.4 cm
\end{figure}

\subsection{Center-of-mass separation}

To illustrate the effectiveness of the method, we make a
digression from decays and continuum and discuss a stable
large-scale shell model example of the center-of-mass (CM)
problem. In Fig. \ref{CMO20} we show the strength function of the
dipole operator. The strength function for a state
$|\lambda\rangle $ is defined as
\begin{equation}
F_\lambda(E)=\langle \lambda | \delta(E-H) |\lambda
\rangle=-\frac{1}{\pi}\, {\rm Im}\, \langle\lambda |G(E)
|\lambda\rangle.                          \label{15}
\end{equation}
For the Hamiltonian $H$ here, we consider the full $s-p-sd-pf$
shell model space with positive parity states restricted to the
$sd$ shell, while the negative parity states include all
one-particle-hole excitations from the $sd$ shell. The two-body
interaction is chosen as WBP \cite{brown01}. A Lawson technique
is used to address the CM problem with an artificial
CM vibration Hamiltonian included into $H$ with a
large scaling factor. As a result, all states that correspond to
the CM excitations appear at high energy, around 100
MeV in our example. The strength function of the dipole operator
${D}=\sum_a e_a r_a$ is considered, where $e_a$ is the effective
charge of a particle $a$. In Fig. \ref{CMO20} the dipole strength
for excitations from the $0^+$ ground state of $^{20}$O is
plotted, namely Eq. (\ref{15}) is evaluated with
$|\lambda_D\rangle=D |{\rm g.s}\rangle.$ Depending on the choice
of the effective charges for protons and neutrons, the operator
$D$ can be changed from the pure CM operator to the isovector
operator containing no CM component. The change in strength of CM
states is shown in Fig. \ref{CMO20}.

\section{Helium isotopes\label{he}}

We conclude with a realistic example of CoSMo application to the chain of
helium isotopes $^4$He to $^{10}$He that serves as an
illustration of all techniques combined. The internal space of
this simplified model contains two single-particle levels
$p_{3/2}$ and $p_{1/2}$ on top of the $\alpha$-particle core. The
sensitivity of the decay amplitudes $A_{1}^{c}$ to the 
location of thresholds and to the parent-daughter structural relations
lead to the necessity of considering the
entire isotope chain. The effective shell model interactions were
taken from \cite{cohen65,stevenson88}. These interactions are
experimentally adjusted; thus it is assumed that the Hermitian
renormalizations due to virtual particle excitations into
continuum, $\Delta(E)$, Eq. (\ref{3}), are already implicitly
included; the energy dependence of $\Delta(E)$ is
neglected. The diagonalization of the Hermitian many-body
Hamiltonian within this valence space provides a conventional
shell model solution.

{\sl One-body} decays are accounted for in the model through the
single-particle decay amplitudes defined as
\begin{equation}
A^c_1(E)= a_j(\epsilon)\,\, \langle 1; N|b^\dagger_j|\alpha;N-1\rangle.
                                                   \label{16}
\end{equation}
These amplitudes correspond to a single particle amplitude $a_j$
of the decay leaving a residual $N-1$ nucleon state $\alpha$,
while the remaining nucleons can be seen as spectators. The
amplitude $a_j$ as a function of energy is determined with the use
of the Woods-Saxon potential that models the single-particle
interaction between bound and continuum states. The parameters of
the potential are adjusted in order to adequately represent the
$^4$He+$n$ scattering.

The {\sl two-body} decays can be separated into
sequential and direct ones. The {\sl sequential} decays represent
higher order processes generated by the same single-particle
mechanism modeled here by the Woods-Saxon potential. The {\sl
direct} decay requires introduction of new parameters describing
instantaneous removal of an interacting pair. The model includes
simplest two-body terms, see further discussion in
\cite{volya_PRC67}.

\begin{figure}
\begin{center}
\includegraphics[width=3.5 in]{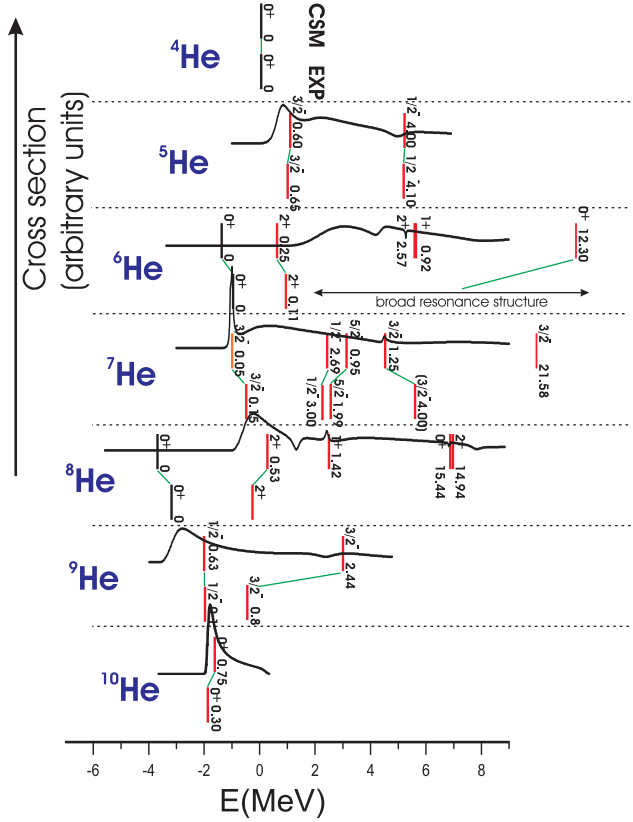}
\end{center}
\caption{(Color online) CoSMo results for He isotopes. The states
in the chain of isotopes starting from $^4$He (top) up to
$^{10}$He (bottom) are shown as a function of the energy relative
to $^4$He. The horizontal dotted lines separate isotopes. For each
case, the states from CoSMo are shown above experimentally
observed states. The decay width (in units of MeV) along with spin
and parity is shown for each state. The solid lines above CoSMo
states show the elastic neutron scattering cross section from the
spin polarized state of $N-1$ isotope 
in the magnetic substate with $M=0$ (even) or
$M=1/2$ (odd mass) quantum number.
\label{he2}} \vspace{-0.2 cm}
\end{figure}

In Fig.~\ref{he2} the results of calculations are shown and
compared to the experimental data
\cite{nndc,rogachev04,korsheninnikov99}. The resonance states
computed according to the Breit-Wigner definition are shown with
discrete lines labeled with spin, parity and decay width. 
The same Fig.~\ref{he2} contains a separate cross section
calculation which implements techniques discussed in Secs.
\ref{unit} and \ref{cheb}. The reaction calculation is performed 
with the use
of the same Hamiltonian (\ref{1}) and thus provides an important
complementary picture to the resonant structure. The cross section
shown is that for elastic neutron scattering off the ground state
of the $N-1$ nucleus.

The model, in agreement with experiment, predicts ground states of
$^{4,6,8}$He to be particle-bound. The energies of these states by
construction of the model exactly agree with the prediction of the
traditional shell model. The states in the continuum are
approached from two perspectives, via the solution of Eq.
(\ref{5}) under the Breit-Wigner resonance condition and by
directly plotting the cross section curve. The resonance centroids
are shown with discrete lines with corresponding widths indicated.
The continuum coupling changes the structure of internal states
leading to resonant energies being in general different from those
from the shell model prediction. The resonant patterns
indirectly reveal information about structure of the states and
dominant decay modes. The results for $^7$He isotope 
agree with recent experiments \cite{rogachev04,korsheninnikov99}.
Our results support the ``unusual structure" of the $5/2^-$ state
identified by \cite{korsheninnikov99}. Due to its relatively high
spin, this state, unlike the neighboring $1/2^-$ state, decays
mainly to the $2^+$ excited state in $^6$He.

The comparison of the cross section curve with discrete resonances
provides a transparent picture revealing both usefulness and
limitations of the resonant parameterization approach. The
scattering cross sections start from thresholds set here by the
ground state of the previous ($N-1$) nucleus. The cross sections
at sharp resonances, such as the ground state of $^7$He, agree
well with the resonance parameterization. However, generally the
cross section curves are not symmetric and do not show a simple,
Gaussian or Lorentzian, shape. The shape of low-lying states with
widths big enough to reach threshold is particularly influenced.
The association of the cross section peaks with the location of
the resonance is ambiguous. A remarkable example is the case of
$1/2^-$ state in $^7$He. The Breit-Wigner approach predicts an
almost 3 MeV wide resonance at 2.3 MeV of excitation energy. The
cross section curve, however, is only weakly influenced by such a
deep pole and peaks near low energies reflecting primarily a
proximity of threshold. This comparison of cross section and
resonance parameterization may shed light onto the experimental
controversy discussed in
\cite{rogachev04,korsheninnikov99,wuosmaa05,meister02}.

\section*{Acknowledgments}

The author thanks V. Zelevinsky for collaboration. The  
support from the U.S. Department of
Energy, grant DE-FG02-92ER40750, and National Science Foundation,
grants PHY-0070911 and PHY-0244453 is acknowledged. 
Useful discussions with B.A. Brown and G. Rogachev are appreciated.


\end{document}